\documentstyle [12pt] {article}
\advance\voffset by -1.1in
\advance\hoffset by -0.3in
\input amssym.def
\input amssym
\def \bbR{{\Bbb R}}

\newcommand{\subsect}[1]{\setcounter{equation}{0}\subsection{#1}}
\def\rref#1{(\ref{#1})}
\newcommand{\beq}{\begin{equation}}
\newcommand{\eeq}{\end{equation}}
\newcommand{\beqa}{\begin{eqnarray}}
\newcommand{\eeqa}{\end{eqnarray}}
\renewcommand{\thesubsection}{\arabic{subsection}}
\renewcommand{\theequation}{\arabic{subsection}.\arabic{equation}}
\newcommand{\nn}{\nonumber}
\def\vev#1{{\langle #1  \rangle}}
\def\mod#1{{| #1  |}}
\def \de{\delta}

\def \Ga{\Gamma}

\def \pr{\partial}
\def \bx{{\bf x}}

\def \0{{\bf 0}}

\def \vep{\varepsilon}
\def \half{{\textstyle {1 \over 2}}}

\def \hlf{{1\over 2}}
\def \txt{ \textstyle}
\def \trho{{\tilde \rho}}

\def \txi{{\tilde {\xi}}}
\def \thf{{\tilde {\hat {f}}}}
\def \thg{{\tilde {\hat {g}}}}
\def \thh{{\tilde {\hat {h}}}}

\def \I{{\cal I}}
\def \O{{\cal O}}
\def \L{{\cal L}}
\def \R{{\cal R}}
\def \V{{\cal V}}

\def \d{{\rm d}}

\def \balpha{{\bar {\alpha}}}

\def \hh {{\hat {h}}}

\def \brs {{\bar s}}
\def \brG {{\bar G}}

\def \brts{{\bar{\tilde{s}}}}
\def \hg{{\hat{g}}}
\def \hf{{\hat{f}}}

\def \br{{\bf r}}
\def \ts{{\tilde s}}

\def \tX{{\tilde X}}

\begin{document}
%%%%%%%%%

\rightline{UBC/TP-95-008}
\rightline{hep-th/9507028}
\vskip 2truecm
\begin{center}
{\large \bf Integral Transforms for Conformal Field Theories
with a Boundary}
\end{center}
\vskip 1.5 true cm
\centerline {D.M. McAvity}
\vskip 10pt
\centerline {\small Department of Physics, University of British Columbia}
\centerline {\small 6224 Agricultural Rd, Vancouver, BC, V6T 1Z2, CANADA}
\centerline {\small email: dmm@physics.ubc.ca}
\vskip 50pt
\begin{quote}
\small
A new method is developed for solving the conformally invariant
integrals that arise in conformal field theories with a boundary.
The presence of a boundary makes previous techniques for theories
without a boundary less suitable. The method makes essential use
of an invertible integral transform, related to the radon transform,
involving integration over planes parallel to the boundary. For
successful application of this method several nontrivial
hypergeometric function relations are also derived.
\end{quote}

\vskip 20pt
\subsect{\bf Introduction}
\label{intro}
\vskip 5pt
At a critical point most statistical mechanical systems are not
only scale invariant but are also conformally
invariant~\cite{cardy:NPB240,wegner:domb}. This principle
has profound implications for calculations of the correlation functions,
critical exponents and universal amplitudes of such
systems~\cite{pg:houches}.
In two dimensions, where the conformal group is infinite dimensional,
multipoint correlation functions are more strongly
constrained then in dimension $d>2$, where the conformal group
is finite. However, consideration of $d>2$ is also important,
particularly in the statistical mechanical context when $d=3$.
In the case of general $d$ conformal invariance still provides
quite powerful constraints. For example, in the infinite
geometry $\bbR^d$ the forms of the two and three point
functions of scalar fields in a conformal field theory are determined
exactly (up to normalisation) by the restrictions of conformal
invariance.

Cardy has shown how to generalise the principle of
conformal invariance to the case of the semi-infinite geometry
$\bbR^d_+$, so that surface critical phenomena can be probed using
these techniques~\cite{cardy:NPB240,cardy:domb}.
In $\bbR^d_+$ it is only appropriate to have
conformal invariance under conformal transformation which leave
the boundary fixed. In this case the restrictions on the form of
correlations functions are not as strong. In particular the
form of the two point
function of a scalar field in $\bbR^d_+$ is restricted by conformal
invariance only up to some function of a single  conformally invariant
variable~\cite{cardy:NPB240}. This function must be then be determined
for the particular theory under consideration.

In this paper we outline a powerful method, which makes essential
use of conformal invariance, for calculating the two point functions
of scalar, vector and tensor fields of conformal field theories in
the semi-infinite space $\bbR^d_+$. In particular we give a
prescription for treating the conformally invariant integrals
that arise in a diagrammatic expansion of the theory. Techniques for
handling such integrals have been developed for the infinite space
$\bbR^d$, and have proven to be very
useful~\cite{peliti:LNC,symanzik:LNC}. However these techniques
do not extend to $\bbR^d_+$ and so this alternative technique is
developed.

\subsect{\bf Conformal Invariance}
\label{conformal}

A transformation of coordinates $x_\mu \to x_\mu^g(x)$  is a
conformal transformation if it leaves the line
element unchanged up to a local scale factor $\Omega(x)$. That is
\beq
\d x^g_\mu \d x^g_\mu =\Omega(x)^{-2}\d x_\mu \d x_\mu \; .
\eeq

For the discussion of two point functions of fields in a conformal
field theory we need to consider the effect of conformal
transformations on these fields. If a field $\O(x)$
transforms under the conformal group as
\beq
\O(x) \to \O^g(x^g)=\Omega(x)^\eta \O(x)\; ,
\eeq
for some $\eta$, then $\O(x)$ is said to be a quasi primary scalar field
with scale dimension $\eta$.
A quasi primary vector field $\V_\mu(x)$ with scale dimension $\eta$
is one which transforms like
\beq
\V(x) \to \V_\mu^g(x^g)=\Omega(x)^\eta\R_{\mu\alpha}(x)\V_\alpha(x)\; ,
\eeq
where $\R_{\mu\alpha}(x)=\Omega(x)\pr x^g_\mu /\pr x_\alpha$. The
transformation for quasi primary tensor fields follows analogously.
We will restrict our attention to quasi primary
fields in this paper.

In the semi-infinite space $\bbR^d_+$ we define coordinates
 $x_\mu=(y,\bx)$ where $y$
measures the perpendicular distance from the boundary, and $\bx_i$
are coordinates in the $d-1$ dimensional hyperplanes parallel to
the boundary. The two point
functions  of scalar operators are restricted by translational
and rotational invariance in planes parallel to the boundary to be
\beq
\vev{\O_1(x)\O_2(x')}=G(y,y',\mod {\bx-\bx'})\; ,
\eeq
and scale invariance further restricts the form of $G$ to depend
on two independent scale invariant variables $s^2/ y^2$ and
$s^2/y'{}^2$, where $s^2=(x-x')^2$. This situation should
be contrasted with the case of infinite space where it is not possible
to construct a variable from two points which is invariant under all of
scale, translational and rotational transformations.

For two points in $\bbR^d_+$ conformal invariance provides further
restrictions. Under conformal transformations which leave the
boundary fixed
\beq
s^2 \to {s^2 \over \Omega(x)\Omega(x')}\, , \quad \quad
y \to {y\over \Omega(x)} \, , \quad \quad  y' \to {y' \over
\Omega(x')} \, ,
\label{eq:confy}
\eeq
so that only one independent conformally invariant variable can
be constructed from two points
\beq
\xi={s^2 \over 4yy'} \, \quad \quad {\rm or} \quad \quad
v^2={s^2 \over \brs^2}={\xi\over 1+\xi} \, ,
\eeq
where $\brs^2=(\bx-\bx')^2 +(y+y')^2$ is the square of the distance
along the path between $x$ and the image point of $x'$ .

As a consequence, the correlation function of two quasi primary scalar
fields may be written as
\beq
\vev {\O_1(x)\O_2(x')} = {1\over (2y)^{\eta_1}}{1\over (2y')^{\eta_2}}
f(\xi)\, ,
\eeq
for some arbitrary function $f(\xi)\;$\footnote{The $\xi \to 0$ and
$\xi \to \infty$ limiting behaviour of this function is fixed by
the Operator Product and Boundary Operator
Expansions~\cite{mca-osb:sur2}.}

As an example we consider free scalar field theory, where the field
$\phi(x)$ satisfies Dirichlet or Neumann boundary conditions at $y=0$.
Then by the method of images the Green's function is simply
\beq
\vev{\phi(x)\phi(x')}=G_\phi(x,x')=A \Big ({1\over s^{d-2}} \pm
{1\over \brs^{d-2}} \Big )={A\over (4yy')^{\eta_\phi}}f_\phi(\xi)
 \, ,\label{eq:Gfree}
\eeq
where
\beq
A={1\over (d-2) S_d}\, , \quad \quad \eta_\phi=\half d-1 \, , \quad
\quad f_\phi(\xi)= \xi^{-\eta_\phi} \pm (1+\xi)^{-\eta_\phi} \; .
\eeq
In the above expression  the upper (lower) sign corresponds to
Neumann (Dirichlet) boundary conditions and the factor
$S_d=2\pi^{\hlf d}/ \Ga(\half d)$ is the area of a unit hypersphere in
$d$ dimensions.

In~\cite{mca-osb:sur2}, henceforth referred to as $I$,
the form of the
two point functions of scalar, vector and
tensor fields was worked out in detail for the $O(N)$ sigma model
in both the $\vep$ and  large $N$ expansions. These calculation were
significantly simplified by the use of a new technique to solve the
conformally invariant integrals on $\bbR^d_+$ that naturally arise.
In the next section this technique is discussed in detail.

\subsect{Parallel Transform Method}
\label{method}

We consider integrals of the form
\beqa
f(\xi) & = & \int_0^\infty\!\! \d z\int \! \d^{d-1}\br \,
{1\over (2z)^d} f_1(\txi) f_2(\txi')\; , \label{eq:integ} \\
 && \txi\, =\, {(x-r)^2\over 4yz}
\quad\quad \txi' \,=\, {(x'-r)^2\over 4y'z} \; ,
\quad\quad r\,=\,(z,\br) \; ,\nn
\eeqa
where conformal invariance restricts the form of the integral to be a
function of $\xi$ only. This follows because under conformal
transformations which leave the boundary fixed,  the
integration measure transforms like $\d^dx \to \Omega(x)^{-d}\d^dx$
and the factor $1/(2z)^d \to \Omega(x)^d/(2z)^d$ so the
local scaling factor cancels.

Given functions $f_1$ and $f_2$ we may solve
integrals of this type indirectly by first integrating $f(\xi)$ over
hyperplanes parallel to the boundary\footnote{This is related to the
Radon transformation of $f(x)$~\cite{gelfand}, which is defined as the
integral of
$f(x)$ over all possible hyperplanes in $\bbR^d$. Here we consider
integrals over the subset of hyperplanes in $\bbR^d_+$  which are
parallel to the boundary.}
\beq
 \int \! \d^{d-1} \bx \, f(\xi) = (4yy')^{\lambda}
{\hat f}(\rho) \; , \quad\quad \rho = {(y-y')^2\over 4yy'} \; ,
\quad\quad \lambda = \half (d-1) \; ,
\label{eq:hf}\eeq
which defines the function $\hf(\rho)$ to be
\beq
{\hat f}(\rho) = {\pi^\lambda \over \Gamma (\lambda)}
\int_0^\infty \!\!  \d u \, u^{\lambda -1}f(u+\rho) \; .
\label{eq:trans}\eeq
The crucial point is that this defines an integral transform
$f \to \hf$  which is invertible.
Thus $f(\xi)$ can be retrieved from $\hf(\rho)$ via
\beq
 f(\xi) = {1\over \pi^\lambda \Gamma (-\lambda)} \int_0^\infty \!\!
\d \rho \, \rho^{-\lambda - 1}{\hat f}(\rho + \xi) \; .
\label{eq:inver}\eeq
The integral in the above formula is actually singular for values of
$\lambda$ that we consider here, but the inversion formula may still be
defined by analytic continuation in $\lambda$ from ${\sl Re}(\lambda)<0$.
To verify that the transformation \rref{eq:trans} is compatible with the
inversion formula \rref{eq:inver} it is sufficient to make use of the
following relation involving generalised functions
\beq
\int \! \d u \, (\rho -u)_+^{\mu-1} u_+^{\lambda-1} = B(\mu,\lambda)\,
\rho_+^{\mu+\lambda-1} \sim \Gamma (-\lambda) \Gamma (\lambda) \de (\rho)
\ \ \hbox{as} \ \ \mu\to -\lambda \; .
\eeq
For the case $d=3$ when $\lambda=1$ we use
\beq
{\rho_+^{-\lambda -1}\over \Gamma(-\lambda)} \sim  \de'(\rho) \quad
 \hbox{as} \quad \lambda \to 1 \, ,
\eeq
to reduce the inversion formula~\rref{eq:inver} to the simple form
\beq
 f(\xi) = - {1\over \pi} \, {\hat f}'(\xi) \, .
\eeq

Now that this parallel transform has been defined it is possible to
derive an integral relation for the transformed functions by
integrating  $f(\xi)$ in~\rref{eq:integ} with respect to $\bx$ so that
\beq
\hf(\rho)= \int_0^\infty\!\! \d z \, {1\over 2z}
\hf_1(\trho)\hf_2(\trho')\, \quad \quad {\trho} =
{(y-z)^2\over 4yz}   \; , \quad {\trho'} =
{(y'-z)^2\over 4y'z} \; .
\label{eq:tran-int}\eeq
In order to solve integrals of this type we first change variables
 $z=e^{2\theta}$, $y=e^{2\theta_1}$ and $y'=e^{2\theta_2}$ so that
equation \rref{eq:tran-int} becomes
\beq
\hf \Big (\sinh^2(\theta_1-\theta_2)\Big ) =
\int_{-\infty}^\infty \!\! \d\theta \,\hf_1 \left
(\sinh^2(\theta-\theta_1)
\right )\hf_2 \left (\sinh^2(\theta-\theta_2)\right )\, .
\label{eq:theta}\eeq
Now by taking the Fourier transform
\beq
{\tilde {\hf}}(k)= \int_{-\infty}^\infty \!\! \d\theta\, e^{ik\theta}
\hf(\sinh^2{\theta}) \; ,
\label{eq:four}
\eeq
then by the convolution theorem the transformed integral relation
\rref{eq:tran-int} becomes
\beq
\thf(k)=\thf_1(k)\thf_2(k) \; .
\label{eq:four-int}
\eeq
Thus we may solve integrals of the general type given
in~\rref{eq:integ} by this double integral transform method provided
that it is possible to make the transforms $f_i(\xi) \to \hf_i(\rho) \to
\thf_i(k)$ for both the functions $f_1$ and $f_2$ {\em and}
that the subsequent inverse transforms of the resulting function
$\thf(k)$ can be made. Of course the form of the functions $f_1$ and
$f_2$ are crucial in order for this procedure to be successfully
undertaken. For the typical cases which arise in the diagrammatic
expansion of a conformal field theory this method has proven to be
very successful, although the intermediate steps often involve
nontrivial manipulations of hypergeometric functions. In the next
section several examples which are likely to occur in calculations in
conformal field theory are given to illustrate the method, and provide
a table of transforms for future reference.

\subsect{Illustration of the Method}
\label{illustrate}

For application to the calculation of two point functions in a
conformal field theory we may use this method to solve the integrals
over products of propagators that occur in a diagrammatic expansion
of the theory. Therefore, by considering, for example, the Green's
function of the free scalar field given in~\rref{eq:Gfree}
we wish to solve integrals of the following type
\beq
I = \int_0^\infty\!\! \d z \int \! \d^{d-1}\br \,{1\over (2z)^\beta}
{1\over \big (\ts^{2}\big )^{\alpha} \big (\brts{}^{2}\big )^{\balpha}
\big ({\ts'^2}\big )^{\alpha'} \big ({\brts{}'^2}\big)^{\balpha'}} \,
 \; , \label{eq:I}
\eeq
with
\beqa
&&\ts^2\; =\; (\bx-\br)^2+(y-z)^2\; , \quad \quad \ \
\brts{}^2\; =\;(\bx-\br)^2+(y+z)^2\; , \nn \\
&&\ts{}'^2 \; =\;(\bx'-\br)^2+(y'-z)^2\; ,\quad \quad
\brts{}'^2 \; =\;(\bx'-\br)^2+(y'+z)^2\; . \nn
\eeqa
For conformal invariance, following~\rref{eq:confy}, we must also require
\beq
\alpha+\balpha+\alpha'+\balpha'+\beta=d \; .
\eeq
This integral may be readily cast into the general
form~\rref{eq:integ}, for which we should then take
\beq
f_1(\txi) ={1\over(2y)^{\alpha+\balpha}}{1\over \txi^\alpha
(1+\txi)^\balpha} \; , \quad \quad \quad
f_2(\txi') = {1\over(2y')^{\alpha'+\balpha'}}{1\over \txi'{}^{\alpha'}
(1+\xi')^{\balpha'}} \; .
\eeq
Later in this section we will consider the more general integrals that
arise in the discussion of the large $N$ expansion of the $O(N)$ sigma
model, where the propagator for the auxiliary field $\lambda$ has
a more  complicated functional form.

To solve the integral~\rref{eq:I} using the method of
section~\ref{method}
we  first take the sequence of transforms $f \to \hf \to
\thf$ as defined in~\rref{eq:trans} and~\rref{eq:four}
for functions of the form $f_i(\xi)$ above.
For simplicity we take
\beq
f(\xi) = {1\over \xi^\alpha (1+\xi)^{\balpha}}
\label{eq:func}\eeq
The first transform $f\to\hf$ follows from standard references
\beqa
\hf(\rho) &=& {\pi^\lambda \over \Gamma (\lambda)}
\int_0^\infty \!\! \d u \, u^{\lambda -1}
{1\over (u+\rho)^\alpha(1+u+\rho)^\balpha} \nn \\
&=& \pi^\lambda{ \Ga(\alpha+\balpha-\lambda) \over \Ga(\alpha+\balpha)}
{1\over (1+\rho)^{\alpha+\balpha-\lambda}}
F\Big (\alpha+\balpha-\lambda,\alpha\,;\,\alpha+\balpha\, ;\,
{1\over 1+\rho} \Big ) \;	 .
\eeqa
The function $F(a,b;c;z)$ is a hypergeometric function whose definition
is given in~\rref{eq:2F1}
For the subsequent transform $\hf\to \thf$ we consider the cases
$\balpha=0,\;\alpha=0,\;\alpha=\balpha$ separately
\beqa
f_{\rm I}(\xi)&=&{1\over \xi^\alpha } \nn \\
\hf_{\rm I}(\rho)&  = &\pi^\lambda {\Ga(\alpha-\lambda)\over
\Ga(\alpha)}{1\over \rho^{\alpha-\lambda}} \label{eq:aa1}  \\
\thf_{\rm I}(k)&=&\pi^{\hlf d}{\Ga(2\alpha-2\lambda)\Ga(1-2\alpha+
2\lambda) \over\Ga(\alpha)\Ga(\half+\alpha-\lambda)} \left [
{\Ga(\alpha-\lambda+{\txt{i\over 2}} k) \over \Ga(1-\alpha+\lambda+{\txt
{i\over 2}} k)} + {\Ga(\alpha-\lambda-{\txt {i\over 2}} k)
\over \Ga(1-\alpha+\lambda-{\txt {i\over 2}}k)} \right ] \nn \\
&&\nn\\
 f_{\rm II}(\xi)&=&{1\over (1+\xi)^\balpha} \nn \\
\hf_{\rm II}(\rho)&  =& \pi^\lambda {\Ga(\balpha-\lambda)\over
\Ga(\balpha)}{1\over (1+\rho)^{\balpha-\lambda}} \label{eq:aa2}\\
\thf_{\rm II}(k)&=&\pi^{\hlf d} {1\over \Ga(\balpha)\Ga(\half+\balpha
-\lambda)}\Ga(\balpha-\lambda +{\txt {i\over 2}} k)\Ga(\balpha-\lambda -
{\txt {i\over 2}}\ k) \nn \\
&&\nn \\
f_{\rm III}(\xi)&=& {
1\over \xi^\alpha (1+\xi)^\alpha} \nn \\
\hf_{\rm III}(\rho)&=&\pi^\lambda {\Ga(2\alpha-\lambda)\over
\Ga(2\alpha)} {1\over(1+\rho)^{2\alpha-\lambda}}F\Big (2\alpha-
\lambda,\alpha\,;\,2\alpha\, ;\, {1\over 1+\rho} \Big ) \label{eq:aa3} \\
\thf_{\rm III}(k)&=&\pi^{\hlf d}4^{\alpha-\hlf d}{\Ga(\half d-\alpha)
\over \Ga(\alpha)} {\Ga(\alpha - \half \lambda- {\txt {i\over 4}} k)
\Ga(\alpha-\half \lambda + {\txt {i\over 4}} k) \over
\Ga(\half+\half \lambda - {\txt {i\over 4}} k)
\Ga(\half+\half \lambda + {\txt {i\over 4}} k) } \nn
\eeqa
There is one other case, a particular combination of two functions of
the type~\rref{eq:func}, which is of interest
\beqa
\hskip -20pt f_{\rm IV}(\xi)&=&
{2\xi+1\over \xi^\alpha (1+\xi)^\alpha}\nn \\
\hskip -20pt\hf_{\rm IV}(\rho)&  = & 2\pi^\lambda
 {\Ga(2\alpha-\lambda-1)\over
\Ga(2\alpha-1)} {1\over (1+\rho)^{2\alpha-\lambda-1}}F\Big (2\alpha-
\lambda-1,\alpha-1\,;\,2\alpha-2\, ;\, {1\over 1+\rho} \Big )
\label{eq:aa4} \\
\hskip -20pt\thf_{\rm IV}(k)&=&\pi^{\hlf d}4^{\alpha-\hlf d}
{\Ga(\half d-\alpha)
\over \Ga(\alpha)} {\Ga(\alpha - \half\lambda-\half -{\txt{i\over4}} k)
\Ga(\alpha-\half \lambda-\half + {\txt {i\over 4}} k) \over
\Ga(\half \lambda - {\txt {i\over 4}} k) \Ga(\half \lambda +
{\txt {i\over 4}} k) }\, .\nn
\eeqa
The last two cases, $f_{\rm III}$  and $f_{\rm IV}$, are important
because the more general case where $\balpha$ differs from $\alpha$
by any integer follows in a straightforward manner from them.
However, the derivation of those two results directly is nontrivial.
The simplest way to verify them is by working
backwards and taking the inverse transforms. A general procedure for
taking the inverse transforms is discussed next.

For application to conformal field theory where we have integrals
of the form~\rref{eq:integ} then the transformed
relation~\rref{eq:four-int}
suggests that we need to take the inverse transform
of products of the functions $\thf_i(k)$ in I to IV.
In all of these cases the dependence of $\thf(k)$ on $k$
is through combinations of Gamma functions. Consequently, by considering
the poles of the Gamma function, the inverse Fourier transform
$\thf \to \hf$ of~\rref{eq:four-int} can be performed by contour
integration. We first consider the following combination
 of Gamma functions
which is appropriate for verifying the transforms of $f_{\rm III}$ and
$f_{\rm IV}$ above
\beq
\thg_{a,b}(k) \equiv  {\Gamma (a-{\txt{i\over 4}}k)
\Gamma (a+{\txt{i\over 4}}k) \over \Gamma (b-{\txt{i\over 4}}k)
\Gamma (b+{\txt{i\over 4}}k) } \, .
\label{eq:thgab}
\eeq
The poles of $\Ga(a-{\txt{i\over 4}}k)$ occur at ${\txt{i\over
4}}k=a+n$ with residue $(-1)^n/n!$ (for $n$ a non negative integer).
Therefore, the inverse transform is obtained as a sum of the residues of
$\thg_{a,b}$, resulting in a series that has hypergeometric form
\beqa
\!\!\!\!\!\!\!\!{\hat g}_{a,b}(\sinh^2 \theta)& = &
{1\over 2\pi}\int \! \d k \, e^{-ik\theta} \, \thg_{a,b}(k) \nn \\
&=& {4\Gamma(2a)\over \Gamma(b-a)\Gamma(b+a)}\,  e^{-4a|\theta|}
F \bigl ( 2a,a-b+1;a+b; e^{-4|\theta|} \bigl ) \, . \label{eq:hgab}  \\
&=& {4\Gamma(2a)\over \Gamma(b-a)\Gamma(b+a)}\,
{1\over (4\cosh^2 \theta)^{2a}}
F \bigl ( 2a,a+b-\half; 2a+2b-1;{1\over \cosh^2\theta}\bigl )\nn \, .
\eeqa
By choosing appropriate values for $a,b$, and noting
that $\cosh^2\theta=1+\rho$ then the Fourier
transformed functions $\thf_{\rm III}$ and $\thf_{\rm IV}$
 follow directly from this result.
To obtain the inverse parallel transform we use
\beq
 {1 \over \Gamma (-\lambda)} \int_0^\infty \!\!\!\!
\d \rho \, \rho^{-\lambda -1} \, {1\over (1+\rho+\xi)^p} =
{\Gamma(p+\lambda)\over\Gamma(p)}{1\over (1+\xi)^{p+\lambda}} \, ,
\eeq
with $p=2a+n$, in the last line of~\rref{eq:hgab}
so that
\beq
g_{a,b}(\xi)= {\Gamma(2a+\lambda)\over 4^{2a-1}\pi^\lambda\Gamma(b-a)
\Gamma(b+a)}\,{1\over(1+ \xi)^{2a+\lambda}}\,
F \Big ( 2a+\lambda,a+b-\half; 2a+2b-1;{1\over 1+\xi}\Big)\; .
\label{eq:gab}\eeq
Now, with the appropriate choice of $a,b$, we can use this result
to verify the parallel transforms $\hf_{\rm III}$ and $\hf_{\rm IV}$
in equations~\rref{eq:aa3} and~\rref{eq:aa4}.

In order to solve the integrals of the type~\rref{eq:integ} we must
find the inverse Fourier transform of products of the functions
$\thf_i(k)$ in I to IV. These may can be simply obtained as
hypergeometric series by contour integration in a similar way to
above above calculation.
The procedure for finding the inverse parallel
transform differs, though, because it is not always possible to make the
simplifying manipulation of the hypergeometric function that is
made in~\rref{eq:hgab}. This is because
the hypergeometric series is often of higher order. However, a
procedure for taking the inverse transform $\hf \to
f$ which bypasses this step is derived in the appendix. This procedure
makes essential use of a special property of the hypergeometric
series which arises on taking the inverse Fourier transform, that is
due the symmetry $\thg(k)=\thg(-k)$. After taking the inverse Fourier
transform of products of the functions in I to IV, we obtain
a hypergeometric series with one of the two following forms
\beqa
\hg(\sinh^2\theta)&= & e^{-4a|\theta|} {}_{q+1}F_q \bigl
(2a,b_1,\cdots b_q; c_1, \cdots c_q ; e^{-4|\theta|} \bigl )\, ,
\label{eq:hgform}\\
\hh(\sinh^2\theta)&=& e^{-2a|\theta|} {}_{q+1}F_q \bigl
( 2a,b_1,\cdots b_q; c_1, \cdots c_q ; e^{-2|\theta|} \bigl )\, ,
\label{eq:hhform}
\eeqa
where the notation ${}_{q+1}F_q$ refers to a generalised
hypergeometric series which is defined in~\rref{eq:pFq}.
The crucial point is that the parameters $b_i$ and $c_i$ in these
functions are always related by $c_i=1+2a-b_i$.

We now present the inverse transforms of six of the
possible combinations of the functions in I to IV, which have been
obtained using this method. These represent solutions to particular
integrals of the type~\rref{eq:integ}.
First we consider products of the functions $f_{\rm I}$ and $f_{\rm
II}$. In these
cases the inverse Fourier transform results in hypergeometric series
of the form~\rref{eq:hhform} and the inverse parallel transform can
be found via the methods outlined in the appendix. Thus,
using~\rref{eq:hresult}, we obtain
\beqa
\I_{\rm I,I}(\xi) & = & \int_0^\infty\!\! \d z\int \!
\d^{d-1}\br \, {1\over (2z)^d} {1\over\txi^\alpha}{1\over
\txi'{}^{\alpha'}} \nn\\
&=&  \pi^{\hlf d} {\Ga(1+\alpha+\alpha'-d)\Ga(\half d-\alpha-\alpha')
\over \Ga(1-\half d)\Ga(\half d)}\,
F\big (\alpha,\alpha';1+\alpha+\alpha'-\half d; -\xi \big ) \nn \\
&& + \;
\pi^{\hlf d} {\Ga(\alpha+\alpha-\half d)\Ga(\half d -\alpha)
\Ga(\half d-\alpha')\over\Ga(d-\alpha-\alpha')\Ga(\alpha)\Ga(\alpha')}
 {1\over \xi^{\alpha+\alpha'-\hlf d}} \nn \\
&& \hskip 100pt \times \; F\big (\half
d -\alpha,\half d -\alpha',1+\half d-\alpha-\alpha'; -\xi\big ) \; ,
\label{eq:I-I}\\
\I_{\rm I,II}(\xi) & = & \int_0^\infty\!\! \d z\int \!
\d^{d-1}\br \, {1\over (2z)^d} {1\over\txi^\alpha}{1\over
(1+\txi')^{\alpha'}} \nn\\
&=&\pi^{\hlf d} {\Ga(1+\alpha+\alpha'-d)\Ga(\half d-\alpha)\over
\Ga(\half d)\Ga(1+\alpha'-\half d)} F(\alpha,\alpha';\half d; -\xi)\; ,
\label{eq:I-II}\\
\I_{\rm II,II}(\xi) & = & \int_0^\infty\!\! \d z\int \!
\d^{d-1}\br \, {1\over (2z)^d} {1\over(1+\txi)^\alpha}{1\over
(1+\txi')^{\alpha'}} \nn\\
&=&\pi^{\hlf d} {\Ga(1+\alpha+\alpha'-d)\over
\Ga(1+\alpha+\alpha'-\half d)} F(\alpha,\alpha';1+\alpha+\alpha'-
\half d;-\xi)\; . \label{eq:II-II}
\eeqa
In order to bring these results to this form
it is necessary to  use several identities of the
hypergeometric function which can be found in the standard
references~\cite{grad}.

If we take the limit $\alpha+\alpha' \to d$ in these integrals,
which corresponds to $\beta \to 0$ in the original
integral~\rref{eq:I} then the following relation
\beq
{1\over \Big ( (x-x')^2\Big )^{\hlf d - \beta}} \sim {1\over 2\beta}\,
S_d \delta^d (x-x') \ \ \hbox{as} \ \ \beta\to 0 \; ,
\eeq
can be used to show that
\beq
\I_{\rm I,I}+\I_{\rm II,II} =\pi^d  {\Ga(\half d -\alpha)
\Ga(\half d-\alpha') \over \Ga(\alpha)\Ga(\alpha')}  \delta^d (x-x')
\; ,
\eeq
in the limit $\alpha+\alpha' \to d$. This is the expected result when
the range of the integral~\rref{eq:I}, with $\balpha=\balpha'=\beta=0$,
is extended to the infinite space $\bbR^d$. In a similar way it is
possible to show that if $\alpha+\alpha'=d$ then $\I_{\rm I,II}+\I_{\rm
II,I}=0$, where $\I_{\rm II,I}$ is defined by taking
$\alpha\leftrightarrow \alpha'$ in $\I_{\rm I,II}$.

We now evaluate three more conformally
invariant integrals involving combinations of the functions  $f_{\rm
III}$ and
$f_{\rm IV}$. In these cases the inverse Fourier transform results in a
hypergeometric series of the form~\rref{eq:hgform}. One obtains
\beqa
\I_{\rm III,III}(\xi)&=&\int_0^\infty\!\!\d z  \int\! \d^{d-1}\br \,
{1\over (2z)^d} {1\over \txi^\alpha (1+\txi)^\alpha}
{1\over \txi'{}^{\alpha'} (1+\txi')^{\alpha'}} \nn \\
&=&  {\Ga(\half d-\alpha')\Ga(\alpha'-\alpha) \Ga(\alpha+
\alpha'-\lambda) \over \Ga(\half d -\alpha)\Ga(\alpha')
\Ga(\alpha+\half)} \pi^{\hlf d}4^{\alpha'-\hlf d}
{1\over [\xi(1+\xi)]^\alpha} \label{eq:III}\\
&&\times \;{}_3F_2\Big (\alpha,1+\alpha-\half d,
\half d-\alpha'; \alpha+\half, 1+\alpha-\alpha';-{1\over 4\xi (1+\xi)}
\Big) \, + \, \alpha \leftrightarrow \alpha'  \; ,\nn \\
\I_{\rm III,IV}(\xi)&=&\int_0^\infty\!\!\d z  \int \!\d^{d-1}\br \,
{1\over (2z)^d} {1\over \txi^\alpha (1+\txi)^\alpha} {2\txi'+
1\over \txi'{}^{\alpha'} (1+\txi')^{\alpha'}} \nn \\
&=&{\Ga(\half d-\alpha)\Ga(\half d-\alpha')\Ga(\alpha+\alpha'-\half d)
\over\Ga(\alpha)\Ga(\alpha')\Ga(d-\alpha-\alpha')}\pi^{\hlf d}
{1\over[\xi(1+\xi)]^{\alpha+\alpha'-\hlf d}} \label{eq:III-IV} \\
&&\times \;F\Big (\lambda -\alpha,\half d -\alpha'\, ;
d-\alpha-\alpha'\, ;
-4\xi(1+\xi) \Big ) \nn \\
\I_{\rm IV,IV}(\xi)& =&\int_0^\infty\!\!\d z  \int \!\d^{d-1}\br \,
{1\over (2z)^d} {2\txi+1\over \txi^\alpha (1+\txi)^\alpha} {2\txi'+
1\over \txi'{}^{\alpha'} (1+\txi')^{\alpha'}} \nn \\
&=&{\Ga(\half d-\alpha')\Ga(\alpha'-\alpha) \Ga(\alpha+
\alpha'-1-\lambda) \over \Ga(\half d -\alpha)\Ga(\alpha')
\Ga(\alpha-\half)} \pi^{\hlf d}4^{\alpha'-\hlf d}
{2\xi+1\over [\xi(1+\xi)]^\alpha} \label{eq:IV}\\
&&\times\; {}_3F_2\Big (\alpha,1+\alpha-\half d,
\half d-\alpha'; \alpha-\half, 1+\alpha-\alpha';- {1\over 4\xi (1+\xi)}
\Big) \, + \, \alpha \leftrightarrow \alpha'\; .\nn
\eeqa
To solve for  $\I_{\rm III,III}$ we require the transformed
function $\thf_{\rm III}(k)$  with the result for the inverse
transform of the general case~\rref{eq:g} which is given in the
appendix. For $\I_{\rm IV,IV}$ we use $\thf_{\rm IV}(k)$ with the
inverse transform~\rref{eq:brg}. To obtain $\I_{\rm III,IV}$ in the
form~\rref{eq:III-IV}, we follow a similar procedure to the other two
cases, but also use a relationship between hypergeometric functions with
argument $-z$ and hypergeometric functions with argument $-1/z$ to
simplify the expression.

The solution to the integrals in~\rref{eq:I-I} to~\rref{eq:IV}
all have a pole at $\alpha=d/2$ except for~\rref{eq:II-II}.
This pole arises due to the short distance logarithmic singularity for
$r\sim x$ in each of these integrals when $\alpha=d/2$.

We are now in a position to evaluate integrals of the
type~\rref{eq:integ} with products of more general functions than
those discussed thus far. For example, if we consider the function
$g_{a,b}$ given in~\rref{eq:gab} which was derived from the definition
of $\thg_{a,b}$ in~\rref{eq:thgab}, then since
\beq
\thg_{a,b}(k)\thg_{b,c}(k)=\thg_{a,c}(k)\; ,
\eeq
it follows directly that
\beqa
\int_0^\infty\!\!\d z \int\! \d^{d-1}\br\,{1\over (2z)^d} g_{a,b}(\txi)
g_{b,c}(\txi') & = & g_{a,c}(\xi) \; , \quad\quad\quad\quad\quad
\quad a \ne c \nn \\
&=&(4yy')^{\hlf d}\de^d(x-x') \; , \quad a=c \; .
\eeqa
This is a solution to an integral of the product of two
hypergeometric functions with the special form~\rref{eq:gab}.
This relation is useful in the large $N$ expansion of the $O(N)$ sigma
model with the Ordinary transition,
where the Green's function of the auxiliary field $\lambda$ is
a hypergeometric function of exactly this
type~\cite{mca-osb:sur2,ohno:let1,ohno:prog}.

We may generalise this further by considering the function
\beq
\thg_{ab,c\de}(k)
 \equiv  {\Gamma (a-{\txt{i\over 4}}k)
\Gamma (a+{\txt{i\over 4}}k)  \Gamma (b-{\txt{i\over 4}}k)
\Gamma (b+{\txt{i\over 4}}k)\over \Gamma (c-{\txt{i\over 4}}k)
\Gamma (c+{\txt{i\over 4}}k) \Gamma (\de-{\txt{i\over 4}}k)
\Gamma (\de+{\txt{i\over 4}}k) } \; .
\label{eq:thgabcd}
\eeq
The methods of the appendix can then be used to obtain the inverse
transforms of this function provided $\de=\half \lambda$ or $\de=\half
+ \half \lambda$.
The inverse Fourier transform gives\footnote{This function, is
related to the Meijer's G-function which is defined by the contour
integration of combinations of Gamma functions with arguments of a
particular form~\cite{grad}.}
\beqa
\hg_{ab,c\de}(\sinh^2\theta)&=&{4\Ga(2a)\Ga(b-a)\Ga(b+a) \over
\Ga(c-a)\Ga(c+a)\Ga(\de-a)\Ga(\de+a)}\, e^{-4a|\theta|}
\label{eq:gabcd1}\\
&&\hskip -10pt\times \; {}_4F_3\Big(
2a,b+a,1+a-c,1+a-\de;1+a-b,c+a,\de+a
\, ; e^{-4|\theta|} \Big ) \nn \\
&& + \, a \leftrightarrow b\; .\nn
\eeqa
Subsequently, using~\rref{eq:g} for the case $\de=\half +\half\lambda$
we find
\beqa
\hskip -35pt g_{ab,c\de}(\xi)&=&{1\over 4^{2a-1}\pi^\lambda}
{\Ga(2a+\lambda)\Ga(b-a)\Ga(b+a) \over
\Ga(c-a)\Ga(c+a)\Ga(\half +\half \lambda-a)\Ga(\half +\half \lambda
+a)} {1\over [\xi(1+\xi)]^{a+\hlf\lambda}} \label{eq:gabcd2}\\
&&\times\;{}_3F_2\Big (a+\half\lambda,\half+a-\half\lambda,c-b\, ;
1+a-b,a+c\, ; -{1\over4\xi(1+\xi)} \Big ) \; \nn \\
&& + \, a \leftrightarrow b\; ,\nn
\eeqa
whereas when $\de=\half \lambda$, using~\rref{eq:brg} we obtain
\beqa
g_{ab,c\de}(\xi)&=&{1\over 4^{2a-1}\pi^\lambda}
{\Ga(2a+\lambda)\Ga(b-a)\Ga(b+a) \over
\Ga(c-a)\Ga(c+a)\Ga(\half \lambda-a)\Ga(\half \lambda+a)}
 {\xi+\half\over [\xi(1+\xi)]^{\hlf+ a+\hlf\lambda}} \\
&&\times\;{}_3F_2\Big (\half+a+\half\lambda,1 +a-\half\lambda,c-b\, ;
1+a-b,a+c\, ; -{1\over4\xi(1+\xi)} \Big ) \nn \\
&& + \, a \leftrightarrow b\; .\nn
\eeqa
Thus provided $\de$ is one of $\half \lambda$ or $\half +\half
\lambda$ then $g_{ab,c\de}(\xi)$ can be obtained as ${}_3F_2$
hypergeometric functions. The solutions to the integrals
in~\rref{eq:III}-\rref{eq:IV} represent special
cases of these functions. More generally, integrals
of products of these types of ${}_3F_2$
hypergeometric functions are possible. Since
\beq
\thg(k)_{ab,c\de}\, \thg_{ce,bf}(k)\;=\; \thg(k)_{ae,\de f} \; ,
\eeq
then it follows that
\beq
\int_0^\infty\!\!\d z \int\! \d^{d-1}\br\,{1\over (2z)^d} g_{ab,c\de}
(\txi) g_{ce,bf}(\txi')\; =\; g_{ae,\de f}(\xi) \; ,
\eeq
provided  $\de,f=\half\lambda,\half+\half \lambda$. Similar integral
relations can be derived by considering possible combinations of
$g_{a,b}$ with $g_{ab,c\de}$ with particular choices of the
parameters $a,b,c,\de$.
Integrals such as
these occur in a discussion of the large $N$ expansion of the $O(N)$
sigma model with the Special transition where the Green's
function for the auxiliary field $\lambda$ contains hypergeometric
functions of this type~\cite{mca-osb:sur2,ohno:prog,ohno:let2}.

\subsect{Integrals Involving Spin Factors}
\label{spin}

We now turn our attention to conformally invariant integrals
involving spin factors which occur in the discussion of two point
functions of vector and tensor fields. For this we define fields
$X_\mu$, $\tX_\mu$, with scale dimension
zero, which transform like vectors at the point $x$
under conformal transformations that leave the boundary fixed.
\beq
X_\mu={y\over \xi^\hlf(1+\xi)^\hlf}\pr_\mu \xi \; , \quad \quad \quad
\tX_\mu= {y\over \txi^\hlf(1+\txi)^\hlf}\pr_\mu \txi \; .
\eeq
These  are constructed to be unit vectors so that
$\; X_\mu X_\mu\, =\, \tX_\mu \tX_\mu\,=\,1\, .$

We will use the example of an integral with
one spin factor in the integrand to illustrate the method. Such an
integral would be appropriate for correlation functions involving a
single vector field. We define
\beq
I_\mu=I(\xi)X_\mu=
\int_0^\infty \!\!\d z \int\!\d^{d-1}\br \, {1\over (2z)^d} \tX_\mu
f_1(\txi) f_2(\txi') \; ,
\label{eq:Ivmu}\eeq
which has the functional form $I_\mu=I(\xi)X_\mu$ due to conformal
invariance. To find $I(\xi)$ we use the fact that $X_\mu$ is a unit
vector to obtain
\beq
I(\xi)=\int_0^\infty \!\!\d z \int\!\d^{d-1}\br \, {1\over (2z)^d} (X\cdot\tX)
f_1(\txi) f_2(\txi')\; .
\label{eq:Iv}\eeq
Now, since
\beq
X\cdot \tX={(2\xi+1)(2\txi+1)-(2\txi'+1) \over 4
\Big (\xi(1+\xi)\txi(1+\txi)\Big )^\hlf}\; ,
\eeq
the methods of section~\ref{illustrate} can be used to solve
for $I(\xi)$ in terms of hypergeometric functions.
For example, if we take
\beq
f_1(\xi)={1\over \xi^\alpha (1+\xi)^\alpha} \; , \quad \quad \quad
f_2(\xi)={1\over \xi^{\alpha'} (1+\xi)^{\alpha'} }\; ,
\eeq
then we may use the solution to the integral $\I_{\rm III,IV}$
 in~\rref{eq:III-IV}
to obtain
\beqa
I(\xi)&=&\pi^{\hlf d}
{\Ga(\half+\half d -\alpha)\Ga(\half d-\alpha')\Ga(\half+\alpha+
\alpha'-\half d) \over\Ga(\half +\alpha)\Ga(\alpha')\Ga(\half+ d
-\alpha-\alpha') } {1\over [\xi(1+\xi)]^{\alpha+\alpha'-\hlf d}} \nn\\
&&\times \; F\Big ( d-2\alpha,d -2\alpha';\half
+d-\alpha-\alpha'; -\xi\Big )\; .
\label{eq:Iresult}\eeqa
The case where $\alpha=\half(d-1)$ can be worked out by an alternative
method by noting that\footnote{For this we recall that $-\pr^2
s^{2-d}=(d-2)S_d\de^d (s)\, .$}
\beq
\pr_\mu\left ({1\over (2y)^{d-1}} {\tX_\mu \over
[\txi(1+\txi)]^{\hlf(d-1)}} \right )=S_d\de^d(x-r)\; ,
\eeq
so that from the definition of $I_\mu$ in~\rref{eq:Ivmu},
\beq
\pr_\mu \left ({1\over (2y)^{d-1}} I_{\mu}\right )=S_d{1\over (2y)^d}
{1\over [\xi(1+\xi)]^{\alpha'}}\; .
\label{eq:prImu}\eeq
This result may be rewritten as a differential equation for $I(\xi)$
\beq
{\d \over \d \xi}\left ( [ \xi(1+\xi) ]^{\hlf(d-1)} I(\xi) \right )
= \half S_d {1\over [\xi(1+\xi)]^{1+\alpha'-\hlf d}} \; ,
\eeq
which may be solved to give
\beq
I(\xi) = {S_d \over d-2\alpha'} \, {1\over [\xi(1+\xi)]^{\alpha'-\hlf}}\,
F\Big (1,d-2\alpha'\,;1+\half d - \alpha'\,;-\xi \Big ) \; .
\eeq
The constant of integration is taken to be zero, because otherwise
the presence of such a term would violate~\rref{eq:prImu} by producing
an extra delta function contribution to the RHS.
This solution is in agreement with~\rref{eq:Iresult}. A similar
procedure can be used for integrals involving more spin factors, which
would be appropriate for correlation function involving
the energy momentum tensor or two vector fields for example.
Such integrals are evaluated in $I$ by a slightly different method.

\subsection{Large $N$ Expansion for the $O(N)$ Model}
\label{largeN}

In this section, by way of conclusion,
we demonstrate the use of the of the parallel transform method to
to calculate two point functions in the $1\over N$
expansion of the $O(N)$
non-linear sigma model for the case of semi-infinite geometry.
As usual, the nonlinear constraint on the fields $\phi_\alpha(x)$;
$\phi^2=N$ can be removed by introducing an auxiliary field
$\lambda(x)$ in the Lagrangian via an interaction term $\L_{\rm
I}=\half \lambda \phi^2$. To analyse the two point functions of the
fields $\phi_\alpha$ and $\lambda$ we first define
\beq
\vev{\phi_\alpha(x) \phi_\beta(x')}=G_\phi(x,x')\de_{\alpha\beta}\; ,
\quad \quad \quad \vev{\lambda(x)\lambda(x')}=G_\lambda(x,x')\; .
\eeq
Then, to zeroth order in the $1\over N$ expansion,
these Green's functions satisfy the following relations~\cite{vas:tmp1}
\beqa
\Big ( -\nabla^2+ \vev{\lambda(x)} \Big ) G_\phi (x,x')&=&
\de^d(x-x')\; ,\label{eq:gphi}\\
\int\! \d^d r\, G_\phi^2(x,r) G_\lambda(r,x')&=&- {2 \over N}
\de^d(x-x')\; .\label{eq:glambda}
\eeqa
Both of these relations may be solved by making use of conformal
invariance and using the parallel transform method discussed in
section~\ref{method}. For this we write
\beq
G_\phi(x,x')={1\over (4yy')^{\eta_\phi}} f_\phi(\xi)\; , \quad \quad
\quad  G_\lambda(x,x')={1\over (4yy')^{\eta_\lambda}}
f_\lambda(\xi)\;  . \label{eq:confg}
\eeq
Since $2\eta_\phi+\eta_\lambda=d$ due to conformal
invariance of the integral in~\rref{eq:glambda}, then the zeroth
order result $\eta_\phi=\half d -1$ implies that
 $\eta_\lambda=2$ to this order. Now, with the scaling relation
  $\vev{\lambda(x)}=A_\lambda/4y^2$, it is possible to obtain
 $G_\phi$  as a solution to a differential equation. Alternatively we
can recast~\rref{eq:gphi} into an integral equation so that the method of
parallel transforms can be used to obtain a solution. Writing
\beq
\int\! \d^d r \, H(x,r)G_\phi(r,x')=\de^d(x-x')\; ,\label{eq:intgphi}
\eeq
requires that
\beq
H(x,x')=\Big (-\nabla^2+{A_\lambda\over 4y^2} \Big )\de^d(x-x') \; .
\eeq
The integral of $H(x,x')$ over planes parallel to the boundary may
be written as
\beq
\int \d^{d-1} \bx \, H(x,x') = {1\over (4yy')^{3\over 2}} \hh(y,y')\; ,
\eeq
defining $\hh$ to be
\beq
\hh (e^{2\theta}, e^{2\theta'}) = \Big (- {\d^2\over \d\theta^2}
+ 1 + A_\lambda \Big ) \delta (\theta -\theta')\; .
\eeq
The subsequent Fourier transform of $\hh(e^{2\theta},
e^{2\theta'})$
gives the simple expression
\beq
\thh(k)= k^2+1+A_\lambda \; .
\eeq
We may now solve for $G_\phi$ by first integrating the integral
equation~\rref{eq:intgphi} over planes parallel to the boundary
and then taking the Fourier transform as defined in~\rref{eq:four}.
The resulting equation is
\beq
\thh(k)\thf_\phi(k)=1 \; ,
\eeq
where $\thf_\phi(k)$ is the transform of the function
$f_\phi(\xi)$ defined in~\rref{eq:confg}.
Consequently the desired result is
\beq
\thf_\phi(k)={1\over \thh(k)}={1\over k^2+1+A_\lambda} \; .
\eeq
If we are now express $\thf_\phi(k)$ as
\beq
\thf_\phi(k)={1\over 16} {\Gamma (\mu+ {\txt{i\over
4}}k)\Gamma (\mu-{\txt{i\over 4}}k) \over
\Gamma (1+\mu+{\txt{i\over 4}}k)\Gamma (1+\mu-{\txt{i\over 4}}k)} \; ,
\quad \quad \quad \mu^2={1+A_\lambda \over 16}\; ,
\eeq
then we may use the result~\rref{eq:gab} to obtain the inverse
transform directly:
\beq
f_\phi(\xi)={1\over 4^{1+2\mu}\pi^\lambda }
{\Ga(2\mu+\lambda )\over \Ga(1+2\mu)} {1\over (1+\xi)^{2\mu+\lambda}}
F\Big (2\mu+\lambda, \half + 2 \mu; 1+4\mu; {1\over 1+\xi} \Big ) \;
{}.
\eeq
This general form for $f_\phi(\xi)$ gives the correct large $N$
Green's function $G_\phi(x,x')$ appropriate for both the Ordinary
and Special transitions in the statistical mechanical context where
we should take $\mu=(d-3)/4$ and $\mu=(d-5)/4$
respectively~\cite{bray:jpa}.
Solutions for $G_\lambda(x,x')$ can now be obtained in a similar way
via the integral equation~\rref{eq:glambda}. Results for
$G_\lambda(x,x')$ for both the Ordinary and Special transitions were
calculated with the parallel transform method in $I$ and also
in~\cite{ohno:let1,ohno:prog,ohno:let2} by a different method.
It would be interesting to see if the next order in the $1\over N$
expansion can be obtained using the methods discussed in this paper;
this is the subject of future research.

\vskip 20pt

\leftline{\large \bf  Acknowledgements}
\vskip 5pt
I wish to thank Hugh Osborn for many useful ideas and suggestions.
This research was funded by a Postdoctoral Research Fellowship from
 the National Science and Engineering Research Council of
Canada.
\vskip 5pt

\appendix
\renewcommand{\thesubsection}{Appendix:}
\renewcommand{\theequation}{\Alph{subsection}.\arabic{equation}}
\subsect{Hypergeometric Function Relations}
\label{hyper}
In this appendix we derive some essential hypergeometric function
relations that are need in section~\ref{illustrate}. We start with
the definition of the hypergeometric function
\beq
F(a,b;c;z)\equiv\sum_{n=0}^\infty  {z^n\over n!}
{(a)_n(b)_n\over (c)_n}\, ,
\label{eq:2F1}
\eeq
where  $(a)_n=\Gamma(a+n)/\Gamma(a)$ is the Pochhammer symbol.
There is a natural generalisation of this definition, which is called
a generalised hypergeometric series
\beq
{}_{p}F_q (a_1, \cdots a_p; c_1, \cdots c_q; z) \equiv\sum_{n=0}^\infty
{z^n\over n!} {(a_1)_n \cdots (a_p)_n\over (c_1)_n\cdots (c_q)_n}\, .
\label{eq:pFq}
\eeq
For application to section~\ref{illustrate} we need to consider
the inverse parallel transform of functions of the following
hypergeometric form
\beqa
\hg(\sinh^2\theta)&= & e^{-4a|\theta|} {}_{q+1}F_q \bigl
(2a,b_1,\cdots b_q; c_1, \cdots c_q ; e^{-4|\theta|} \bigl )\, ,
\label{eq:hg}\\
\hh(\sinh^2\theta)&=& e^{-2a|\theta|} {}_{q+1}F_q \bigl
( 2a,b_1,\cdots b_q; c_1, \cdots c_q ; e^{-2|\theta|} \bigl )\, ,
\label{eq:hh}
\eeqa
where $c_i=1+2a-b_i$. To take the inverse transform we express these
results as the sum indicated by~\rref{eq:pFq} and then observe that,
with $\rho=\sinh^2\theta$,
\beq
e^{-2p|{\theta}|}=
\Big(\sqrt{\rho^{\vphantom f}}+\sqrt{1+\rho}\Big)^{-2p}={1\over
4^p(1+\rho)^p}F\Big (p,\half+p;1+2p;{1\over 1+\rho}\Big ) \, ,
\eeq
where $p=2a+2n$ for $\hg$ and $p= a+n$ for $\hh$.
We can now obtain the inverse transform by using
\beq
 {1 \over \Gamma (-\lambda)} \int_0^\infty \!\!\!\!
\d \rho \, \rho^{-\lambda -1} \, {1\over (1+\rho+\xi)^p} =
{\Gamma(p+\lambda)\over\Gamma(p)}{1\over (1+\xi)^{p+\lambda}} \, ,
\eeq
with the result
\beqa
&&\hskip -55pt {\Ga(p)\over \Ga(p+\lambda)\Ga(-\lambda)}
\int_0^\infty \!\! \d\rho \, \rho^{-\lambda-1}
\, {1\over (1+\rho+\xi)^p}F\Big (p,\half+p\, ;1+2p\, ;{1\over 1+\rho+\xi}
\Big ) \nn \\
&&\hskip -35pt =  {1\over (1+\xi)^{p+\lambda}}
F\Big (p+\lambda,\half+p\, ;1+2p\, ;{1\over 1+\xi}\Big )
={1\over \xi^{p+\lambda}}
F\Big (p+\lambda,\half+p\, ;1+2p\, ;-{1\over \xi}\Big )
\label{eq:exp-1} \\
&&\hskip -35pt ={\xi+\half\over\big [\xi(1+\xi)\big]^{\hlf
(p+\lambda+1)}} F\Big (\half(p+\lambda+1),1+\half(p-\lambda);1+p\, ;
-{1\over 4\xi(1+\xi)} \Big )\label{eq:exp-2}  \\
&&\hskip -35pt={1\over \big [ \xi(1+\xi)\big ]^{\hlf (p+\lambda)}}
F\Big (\half(p+\lambda),\half(1+p-\lambda);1+p\, ;-{1\over 4\xi(1+\xi)}
\Big )  \;  . \label{eq:exp-3}
\eeqa
The techniques for finding the inverse transform of
$\hg$ and $\hh$ are related,
except that for $\hh$ it is necessary to use~\rref{eq:exp-1} while
for $\hg$ either of the equivalent
results~\rref{eq:exp-2},\rref{eq:exp-3} may be used. Both of these
equivalent results
are helpful, because two different expressions for $g(\xi)$ can be
derived from them and a nice simplification of these expressions
occurs for different values of the parameters $b_i$. For the purpose of
this discussion we will focus on the inversion of $\hg$ for which we
use~\rref{eq:exp-3} with $p=2a+2n$ to obtain
\beqa
\hskip -55pt g(\xi)&=&{1\over 4^{2a}\pi^\lambda}{\Ga(2a+\lambda)\over
\Ga(2a)}{1\over \big [ \xi(1+\xi)\big ]^{a+\hlf\lambda}}
\sum_{n=0}^\infty  {1\over n!}
{(2a)_n(b_1)_n\cdots(b_q)_n\over  (c_1)_n\cdots(c_q)_n }
{(2a+\lambda)_{2n} \over 4^n(2a)_{2n} }  \\
&& \times\; {1\over \big [ 4\xi(1+\xi)\big ]^n}
F\Big (a+\half\lambda+n,\half+a-\half\lambda+n\, ;1+2a+2n\, ;
-{1\over 4\xi(1+\xi)} \Big ) \, . \nn
\eeqa
By expanding the hypergeometric function,
$g(\xi)$ can be rewritten as
\beq
g(\xi)={1\over 4^{2a}\pi^\lambda}{\Ga(2a+\lambda)\over\Ga(2a)}
{1\over \big [ \xi(1+\xi)\big ]^{a+\hlf\lambda}}
\sum_{N=0}^\infty  G_N {(-1)^N\over [4\xi(1+\xi)]^N} \, ,
\label{eq:g}\eeq
where the coefficient $G_N$ is given by the finite sum
\beq
G_N\!=\!\sum_{n=0}^N  {(-1)^n\over n!(N-n)!}
{(2a)_n(b_1)_n\cdots(b_q)_n\over  (c_1)_n\cdots(c_q)_n }
{(2a+\lambda)_{2n} (a+\half\lambda+n)_{N-n}(\half+a-\half
\lambda+n)_{N-n} \over 4^n(2a)_{2n} (1+2a+2n)_{N-n} } \, .
\eeq
This can be simplified further by using the following identities
for the Pochhammer symbol
\beq
(p)_{2n}=4^n(\half p)_n(\half+\half p)_n \, , \quad \quad
(p+n)_{N-n}={(p)_N\over (p)_n} \, , \quad \quad (p)_{-n}={(-1)^n
\over(1-p)_n} \; .
\eeq
so that $G_N$ becomes
\beqa
&&\hskip -25pt G_N= {(a+\half\lambda)_N(\half+a-\half\lambda)_N
\over N! (1+2a)_N}
{}_{q+4}F_{q+3} \Big
(2a,1+a,b_1,\cdots,b_q,\half+a+\half\lambda,-N;\label{eq:GN}\\
&& \hskip 200pt a,c_1,\cdots,c_q,\half+a-\half\lambda,1+2a+N ; 1
\Big ) \nn
\eeqa
The  finite ${}_{q+4}F_{q+3}$ finite hypergeometric series, with
argument 1, has a
special form because $c_i=1+2a-b_i$. As a consequence, for the
particular case  $q=1$, the resulting  ${}_5F_4$ can be summed
exactly by a special limit of Dougall's theorem which
states that~\cite{slater}
\beq
{}_5F_4(2a,1+a,b,c,-N; a,1+2a-b,1+2a-c,1+2a+N;1) ={(1+2a)_N
(1+2a-b-c)_N\over (1+2a-b)_N(1+2a-c)_N}\; .
\eeq
So for $q=1$ and applying Dougall's theorem we find
\beq
g(\xi)={1\over 4^{2a}\pi^\lambda}{\Ga(2a+\lambda)\over\Ga(2a)}
{1\over \big [ \xi(\xi+1)\big ]^{a+\hlf\lambda}}
{}_2F_1\Big (a+\half\lambda,\half-\half\lambda+a-b;1+2a-b;-{1\over
4\xi(1+\xi)} \Big ) \; .
\eeq
This result and the inverse fourier transform~\rref{eq:hgab}
are sufficient for verifying the integral transforms
in~\rref{eq:aa3} and~\rref{eq:aa4}.

Although a generalisation of Dougall's theorem is not known for
arbitrary $q$,  the coefficient $G_N$
can usually be simplified to give a finite ${}_5F_4$ series through
cancellation of the parameters. However, for this cancellation to
occur it may be necessary in some cases to use  an alternative
formula for $g(\xi)$ which may be derived from~\rref{eq:exp-2}
following a similar procedure. The result is
\beq
g(\xi)={1\over 4^{2a}\pi^\lambda}{\Ga(2a+\lambda)\over\Ga(2a)}
{\xi +\half\over \big [ \xi(1+\xi)\big ]^{\hlf+a+\hlf\lambda}}
\sum_{N=0}^\infty  \brG_N {(-1)^N\over [4\xi(1+\xi)]^N} \; ,
\label{eq:brg}\eeq
where
\beqa
&& \hskip -20pt \brG_N = {(\half+a+\half\lambda)_N(1+a-\half\lambda)_N
\over N! (1+2a)_N}
{}_{q+4}F_{q+3} \Big
(2a,1+a,b_1,\cdots,b_q,a+\half\lambda,-N;\label{eq:brGN}\\
&& \hskip 220pt a,c_1,\cdots,c_q,1+a-\half\lambda,1+2a+N\, ; 1
\Big ) \; .\nn
\eeqa

The calculation of  $h(\xi)$ proceeds in a similar way.
Using~\rref{eq:exp-1} one gets
\beq
h(\xi)={1\over 4^{a}\pi^\lambda}{\Ga(a+\lambda)\over\Ga(a)}
{1 \over \xi^{a+\lambda}}\sum_{N=0}^\infty  H_N {(-1)^N\over \xi^N}\; ,
\label{eq:h}
\eeq
where
\beqa
\hskip -20pt H_N&=& {(a+\lambda)_N(\half +a)_N
\over N! (1+2a)_N}
{}_{q+3}F_{q+2} \Big
(2a,1+a,b_1,\cdots,b_q,-N ;\label{eq:HN}\\
&& \hskip 200pt a,c_1,\cdots,c_q,1+2a+N\, ; 1 \Big ) \; . \nn
\eeqa
Again we recall that $c_i=1+2a-b_i$.
For the case $q=1$ which is relevant for section~\ref{illustrate}
we use a theorem similar to Dougall's,
\beq
{}_4F_3\Big(2a,1+a,b,-N;a,1+2a-b,1+2a+N; 1 \Big)=
{(1+2a)_N(\half+a-b)_N\over (\half+a)_N(1+2a-b)_N}\; ,
\eeq
to obtain
\beq
h(\xi)={1\over 4^{a}\pi^\lambda}{\Ga(a+\lambda)\over\Ga(a)}
{1 \over \xi^{a+\lambda}}F\Big (a+\lambda,\half+a-b;1+2a-b\, ;
-{1\over \xi} \Big )\, .
\label{eq:hresult}
\eeq

%\bibliographystyle{../tex/hep}
%\bibliography{ref}

\end{document}